\documentclass[a4paper,10pt]{article}
\usepackage[utf8x]{inputenc}
\usepackage{rsc}
\usepackage{amsfonts}
\usepackage{amsmath}
\usepackage{amssymb}
\usepackage[american]{babel}
\usepackage[T1]{fontenc}
\usepackage{graphicx}

\newcommand{\eps}{\mathcal{E}}

\title{Weakly nonlinear rheology of transiently crosslinked biopolymer gels}
\author{
  Lars Wolff$^*$ \and Klaus Kroy\thanks{Institute for Theoretical Physics,
University of  Leipzig, Germany}}

\begin{document}

\maketitle

\begin{abstract}

Recent experimental investigations have revealed a non-Maxwellian 
absorption pattern in the rheological spectra of actin gels, which was 
interpreted in terms of transient bonds. Here we examine the 
consequences of reversible crosslinking on the apparent linear spectra 
of biopolymer solutions theoretically. For a schematic model consisting 
of a reversibly crosslinked power-law fluid we obtain a simple analytical 
prediction for the position of the absorption peak, which is backed up by
a numerical evaluation of the inelastic glassy wormlike chain model.  This
establishes bond breaking as a nonlinear non-equilibrium effect that can already
be significant for very small driving amplitudes. Our results may be useful for
inferring binding affinities and reaction rates of biochemical crosslinkers
from rheological measurements of {\it in-vitro} reconstituted cytoskeletal gels.

\end{abstract}

\section{Introduction}

Due to rapid recent progress in the biological sciences an increasingly detailed
understanding of the biochemical response and regulation of cells is
quickly emerging. On the other hand, also the response of cells to mechanical
stimulus has been shown to be of crucial physiological importance
\cite{Discher2005,Yeung2005}. In this research area, sophisticated experimental
techniques are at hand
\cite{Semmrich2007,Trepat2007,Fernandez2008,Krishnan2009,Hoffman2009}
and a rich and exciting phenomenology has been uncovered. Attempts have been
made to organize the mosaic of observational evidence by identifying universal
patterns such as power-law rheology \cite{Fabry2001,Fabry2003,Hoffman2009},
the pivotal role of prestress in cell mechanics and for the physiology of
adhering cells \cite{Wang2002, Sunyer2009, Chowdhury2010}, and the fluidization
response of adhering cells upon transient stretch
\cite{Trepat2007,Krishnan2009,Chen2010}. Despite the advances in the
phenomenological characterization, the underlying physical and molecular
mechanisms are still debated \cite{Gittes1998,Fabry2003,Ingber:2003,Kroy2007}. 

One strategy to unravel the physical foundations of cell mechanics is the
so-called bottom-up approach \cite{Bausch2006,Liu2009}. It aims at
isolating so-called fuctional modules of manageable complexity that can be
reconstituted and studied{\it in vitro}. Examples for the success of this
approach
range from reconstituted biomimetic membrane systems \cite{Loose2009},
over cytoskeletal polymer networks
\cite{Wachsstock1994,Xu1998a,Caspi1998,Gardel2006,Lieleg2010} and motility
assays \cite{Loisel1999,Schaller2010} to conceptual progress in the
understanding of complex {\it in-vivo} functions such as cell adhesion
\cite{Geiger2009}.

It is commonly agreed that the actin cortex is a main contributor to the
mechanical response of tissue cells \cite{Bausch2006,Fernandez2007}. Therefore,
reconstituted actin networks with and without crosslinking molecules are studied
by many groups and cell-like mechanical behavior has indeed been demonstrated
\cite{GardelNakamura:2006,Koenderink2009}. An interesting feature that has been
observed in actin networks crosslinked by a single type of crosslinker is the
emergence of a distinguished frequency scale in the frequency-dependent rheology
\cite{Tharmann2007,Lieleg2008}. A non-Maxwellian relaxation pattern
consisting of a pronounced peak appears in the loss modulus, accompanied by a
shoulder in the storage modulus. The increased dissipation has been attributed
to the breaking of transient bonds and a phenomenological connection between the
zero-force off rate of the crosslinker and the characteristic frequency seen in
rheology has been proposed \cite{Wachsstock1994, Lieleg2008,Lieleg2009}. This
proposition does
not contradict the fact that comparable effects are commonly not reported for
cells, as in cells there are many different types of crosslinkers with a whole
distribution of off rates present. 

It therefore seems that a comprehensive theoretical model accounting for
breaking and reforming of transient bonds could contribute much to the
understanding of the mechanics of biopolymer networks and of cells. It has
been argued that a good model should capture the nonequilibrium character
of cell mechanical processes on top of the polymeric basis of cytoskeletal
elasticity and bridge the length and time scales between the elementary
molecular events and meaningful experimental model systems \cite{Fletcher2009}.
This is a precise characterization of what the following analytical and
numerical analysis aims at. We analyze the problem on two levels. First, we
study analytically the influence of bond-breaking on the nonlinear response of a
schematic model for biopolymer networks, a transiently crosslinked power-law
fluid. We analytically derive an absorption pattern in the frequency-dependent
rheology and provide a prediction for its position as a function of parameters
such as the equilibrium off rate and the relative binding affinity. Second, we
report numerical results obtained with a recently proposed generic model for the
inelastic mechanics of biopolymer networks, the inelastic glassy wormlike chain
\cite{Wolff2010}. We find good qualitative agreement of both theoretical
predictions with each other and with previously published results for
crosslinked actin networks \cite{Lieleg2008,Lieleg2009}.

\section{Models}

\subsection{Model for the bond ensemble}

\label{sec:model_bonds}

Consider an ensemble of non-interacting bonds which have two well-defined
states, the bound state at $x_b$ and the unbound state at $x_u$. 
By convention, the energy vanishes in the force-free bound state, and the
unbound state has energy $U$. To go from one state to the other, a bond has to
pass through a metastable transition state of energy $\mathcal E$ at $x_t$.
External forces $f$ are assumed to adiabatically tilt the energy landscape
\cite{Bell1978a,Evans1997}. This phenomenological assumption can be recovered
from more elaborate theories
\cite{Evans1997,Nguyen-Duong2003,Hummer2003,Dudko2006,Walton2008,Dudko2008,
Tshiprut2008, Freund2009} as an approximation limited to certain experimental
conditions. Beyond that, we do not consider these more elaborate theories as
particularly helpful for our present task, which is to derive a robust
practical analytical prediction as to how the rheology of biopolymer networks
is affected by bond breaking. First, most of them disregard rebinding, which is
expected to be relevant for us, second they cannot be straightforwardly
implemented withour either using a much more elaborate model of the polymer
network or invoking further speculative assumptions, {\it e.g.\ }on the
geometry of and the force transduction through the network. We therefore choose
to stick (for now) to the simple Bell model \cite{Bell1978a}, in which the
rates $k_+$ and $k_-$ for binding and unbinding are given by
\begin{equation}
    k_+(f)=k e^{U-\Delta x_u f},
  \label{eq:kon}
\end{equation}
and
\begin{equation}
    k_-(f)=k e^{\Delta x_b f},
  \label{eq:koff}
\end{equation}
with $\Delta x_{b/u}=\vert x_{b/u}-x_t\vert$, respectively. To simplify the
notation, we measure energies in units of the thermal energy $k_B T$, giving
lengths and moduli the dimensions of inverse force and and force squared,
respectively. The prefactor $k$ corresponds to the equilibrium off rate . In the
framework of Kramers' theory \cite{Kramers1940}, it is proportional to the
microscopic attempt rate and decreases exponentially with the barrier height,
$k\equiv e^{- \mathcal E}\tau_0^{-1}$. More sophisticated theoretical
approaches predict different functional forms of the equilibrium off rate
\cite{Evans1997,Nguyen-Duong2003,Hummer2003,Walton2008,Tshiprut2008,
Freund2009}, also including dependencies of the equilibrium off rate on the
stiffness of the force probe
\cite{Nguyen-Duong2003,Walton2008,Tshiprut2008,Freund2009} and even explicitly
take into account a pulling-induced time-dependent free-energy landscape
\cite{Hummer2003,Dudko2006,Dudko2008,Tshiprut2008,Freund2009}. However, at the
present state we are not interested in the microscopic model details and prefer
to treat $k$ as an independent parameter, unless we have reason not to do so
(further model detrails may {\it e.g.\ }matter when one tries to extract
biochemical data from rheological experiments -- see
section~\ref{sec:results_parameters}).

Noting that the fraction of closed bonds and the fraction of open bonds have to
add to one, the fraction $\nu$ of bonds in the closed state (``bond fraction'')
is given by the first-order kinetic equation
\begin{equation}
    \dot \nu(t)=-(k_+ + k_-)\nu(t)+k_+.
  \label{eq:nu}
\end{equation}

We are interested in the linear response of the bond fraction to a sinusoidal
driving force,
\begin{equation}
    f(t)=\hat f \sin(w_0 t).
  \label{eq:driving_force}
\end{equation}
Expanding the solution of equations~(\ref{eq:kon})-(\ref{eq:driving_force}) with
respect to $\hat f$, truncating at linear order, and neglecting transients in
the long-time limit, $t\gg k(e^{U}+1)$, yields
\begin{equation}
    \nu(t)\approx \nu(f=0)-\hat f\cdot
(\alpha_\nu'(\omega_0)\sin(\omega_0 t)-\alpha_\nu''(\omega_0) \cos(\omega_0 t)),
\end{equation}
with the abbreviations
\begin{equation}
    \alpha_\nu'(\omega)=\frac{(\Delta x_b+\Delta x_u)}{4 \cosh^2(
U/2)}\frac1{1+\left(\frac{\omega}{k(1+e^{ U})}\right)^2}
  \label{eq:bonds_linear_re}
\end{equation}
and
\begin{equation}
    \alpha_\nu''(\omega)=\frac{ (\Delta x_b+\Delta x_u)}{4 \cosh^2(
U/2)}\frac{\frac{\omega}{k(1+e^{U})}}{1+\left(\frac{\omega}{k(1+e^{
U})}\right)^2}.
  \label{eq:bonds_linear_im}
\end{equation}

The functions $\alpha_\nu'(\omega)$ and $\alpha_\nu''(\omega)$ are thus
identified as the linear susceptibilities of the change in bond fraction
$\delta \nu$ to a small sinusoidal driving force. Formally, they have the shape
of the susceptibilities of a Kelvin-Voigt material. Note, however, that at the
current stage, no mechanical interpretation can be made and that the mechanical
response of a material vulnerable to internal bond breaking will not
automatically resemble that of a Kelvin-Voigt material (see sections
\ref{sec:results_power-law} and \ref{sec:results_GWLC}). Note further that the
frequency scale is set by $k\cdot (1+e^{U})$ and not by the equilibrium
off rate $k$ alone, as one na\^ively might have expected.
As an aside, we remark that for our present purpose there is no distinction
between the unbinding and the partial unfolding \cite{Hoffman2007} of the
crosslinking protein, if the unfolded structure is sufficiently long and floppy,
but that there is recent evidence from single-molecule experiments
\cite{Ferrer2008} that bond-breaking could be the more common mechanism for
cytoskeletal networks.

\subsection{Power-law fluid}

\label{sec:model_power-law}

As a qualitative, analytically tractable model for a transiently crosslinked
biopolymer network, we consider a model power-law fluid that displays weak
power-law rheology \cite{Fabry2001} over the frequency range of interest. To
enter bond-breaking, we assume that the linear susceptibility of the power-law
fluid is modulated by the inverse bond fraction, {\it i.e.}
\begin{equation}
    \alpha_S'(\omega)= G_0^{-1} \nu(\omega)^{-1} \omega^{-\delta}
  \label{eq:alphaSprime}
\end{equation}
and
\begin{equation}
    \alpha_S''(\omega)=  \tan\left(\delta\frac{\pi}{2}\right)G_0^{-1}
\nu(\omega)^{-1} 
\omega^{-\delta},
  \label{eq:alphaSdoubleprime}
\end{equation}
with an unspecified prefactor $G_0^{-1}$. 

\begin{figure}[h]
    \begin{center}
        \includegraphics[width=6 cm]{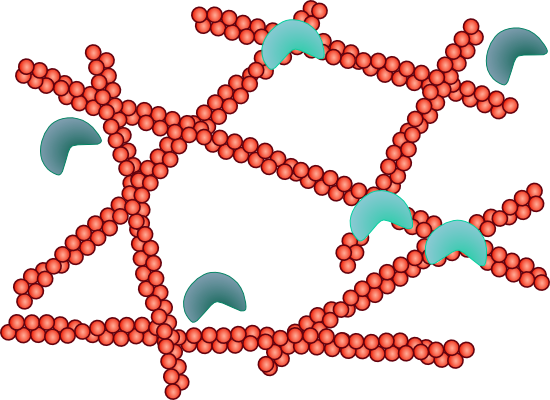}
    \end{center}
  \caption{Sketch of a transiently crosslinked polymer network. In thermal
equilibrium, some polymer-polymer contacts are crosslinked (bright circle
segments), while other crosslinkers are unbound (dark circle segments).
For simplicity, we use a reduced description where direct polymer-polymer
contacts react by effective first-oder kinetics.}
  \label{fig:cartoon}
\end{figure}

\subsection{The inelastic Glassy Wormlike Chain}

\label{sec:model_GWLC}

The glassy wormlike chain (GWLC) model is a phenomenological extension of the
wormlike chain (WLC) model, which is the standard model for individual
semiflexible polymers such as DNA or F-actin. For an introduction to the GWLC,
see Ref.~\citenum{Kroy2007}. A detailed description of the {\em inelastic} GWLC
is given in Ref.~\citenum{Wolff2010}. 

In the weakly bending rod approximation, {\it i.e.\ }if thermal forces do not
bend the polymer too strongly, the WLC can be solved analytically by a mode
decomposition ansatz for the transverse deflections from the straight ground
state. The individual modes with index $n$ decay exponentially with a time
constant $\tau_n$. The GWLC rests on the phenomenological assumption that the
relaxation of modes with wavelength $\lambda_n$ larger than the mean bond
distance $\Lambda$ is slowed down, because $\lambda_n/\Lambda-1$ free energy
barriers have to be overcome for the mode to relax. Formally, this is expressed
by a stretching of the mode relaxation times,
\begin{equation}
  \tau_n \to 	\tilde{\tau}_n=\left\{ \begin{array}{c c} \tau_n &
\lambda_n < \Lambda \\ 
	\tau_n \exp\left[ \eps(\lambda_n/\Lambda -1)\right] & \lambda_n
\geq \Lambda \end{array}\right. .
\end{equation}
In the limit of large free-energy barriers $\mathcal E$, the viscoelastic
response of the GWLC resembles that of a power-law fluid with exponent
\cite{Semmrich2007}
$3\mathcal E^{-1}$.

In the equilibrium GWLC model, the mean distance $\Lambda$ between
polymer-polymer junctions is taken to be a constant $\Lambda_0$, unaffected by
the equilibrium bond fluctuations. The inelastic GWLC keeps track of forced
inelastic bond breaking in the network by dynamically updating $\Lambda$ if the
mean fraction $\nu$ of closed bonds in the network changes:
$\Lambda(t)=\Lambda_0/\nu(t)$. In the simplest version of the inelastic GWLC,
the dynamics of the bond network is approximated by an effective first-order
kinetics and given by the model presented in section~\ref{sec:model_bonds}. See
figure~\ref{fig:cartoon} for an artistic sketch of the physical picture.

\section{Results}

\subsection{Apparent linear response of a transiently crosslinked power-law
fluid}

\label{sec:results_power-law}

To gain some analytical insights, we examine the influence of bond breaking on
the linear response of a schematic power-law fluid, as outlined in
section~\ref{sec:model_power-law},
equations~(\ref{eq:alphaSprime})-(\ref{eq:alphaSdoubleprime}). To this end, we
first assume that both the bonds as well the material in the absence of bond
breaking exhibit a linear response. For the sinusoidal driving force in
equation~(\ref{eq:driving_force}), the displacement response can be written as
\begin{eqnarray}
    x(t)&=&\frac{G_0^{-1} \hat f \omega^{-\delta}}{\nu_0-\hat f\cdot
(\alpha_\nu'(\omega_0)\sin(\omega_0 t)-\alpha_\nu''(\omega_0) \cos(\omega_0
t))}\nonumber \\
  &&\times\left(\sin(\omega_0 t) -\tan(\delta\pi/2 ) \cos(\omega_0 t)\right),
  \label{eq:response}
\end{eqnarray}
where $\nu_0$ is the steady-state bond fraction in the absence of driving. If we
expand the right-hand side to first oder in $\hat f$, we obtain
\begin{equation}
    x(t)\approx\frac{G_0^{-1} \hat f \omega^{-\delta}}{\nu_0}\nonumber
\left(\sin(\omega_0 t) -\tan(\delta\pi/2 ) \cos(\omega_0 t)\right),
\end{equation}
{\it i.e.} the bond breaking does not contribute to the linear response. 

\begin{figure}[h!]
\begin{center}
$\begin{array}{c@{\hspace{3 mm}}c}
\multicolumn{1}{l}{\mbox{\bf\Large a}} &
	\multicolumn{1}{l}{\mbox{\bf\Large  b}} \\ [-0.43cm]\\
  \includegraphics[width=6 cm]{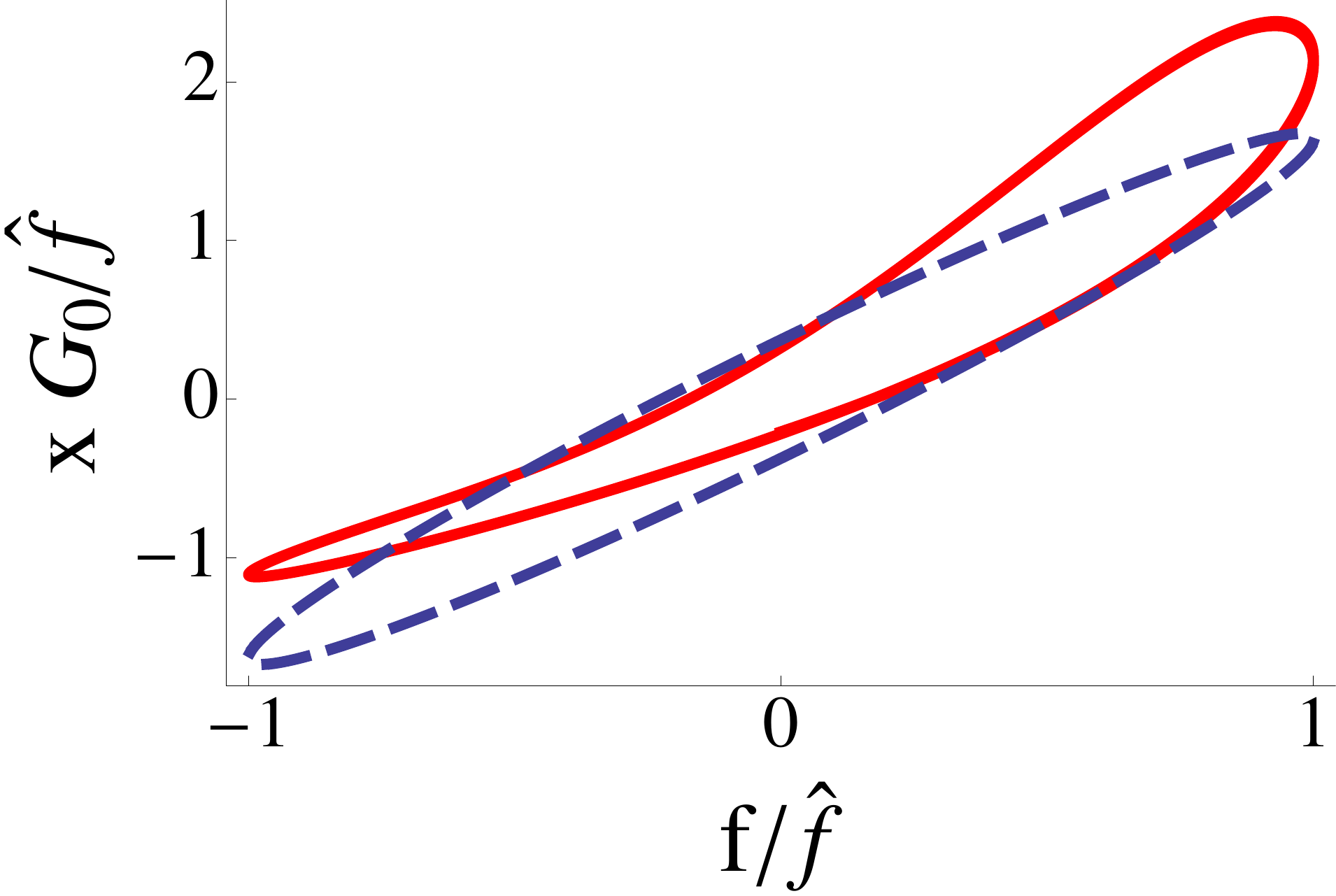} &
	\includegraphics[width=6 cm]{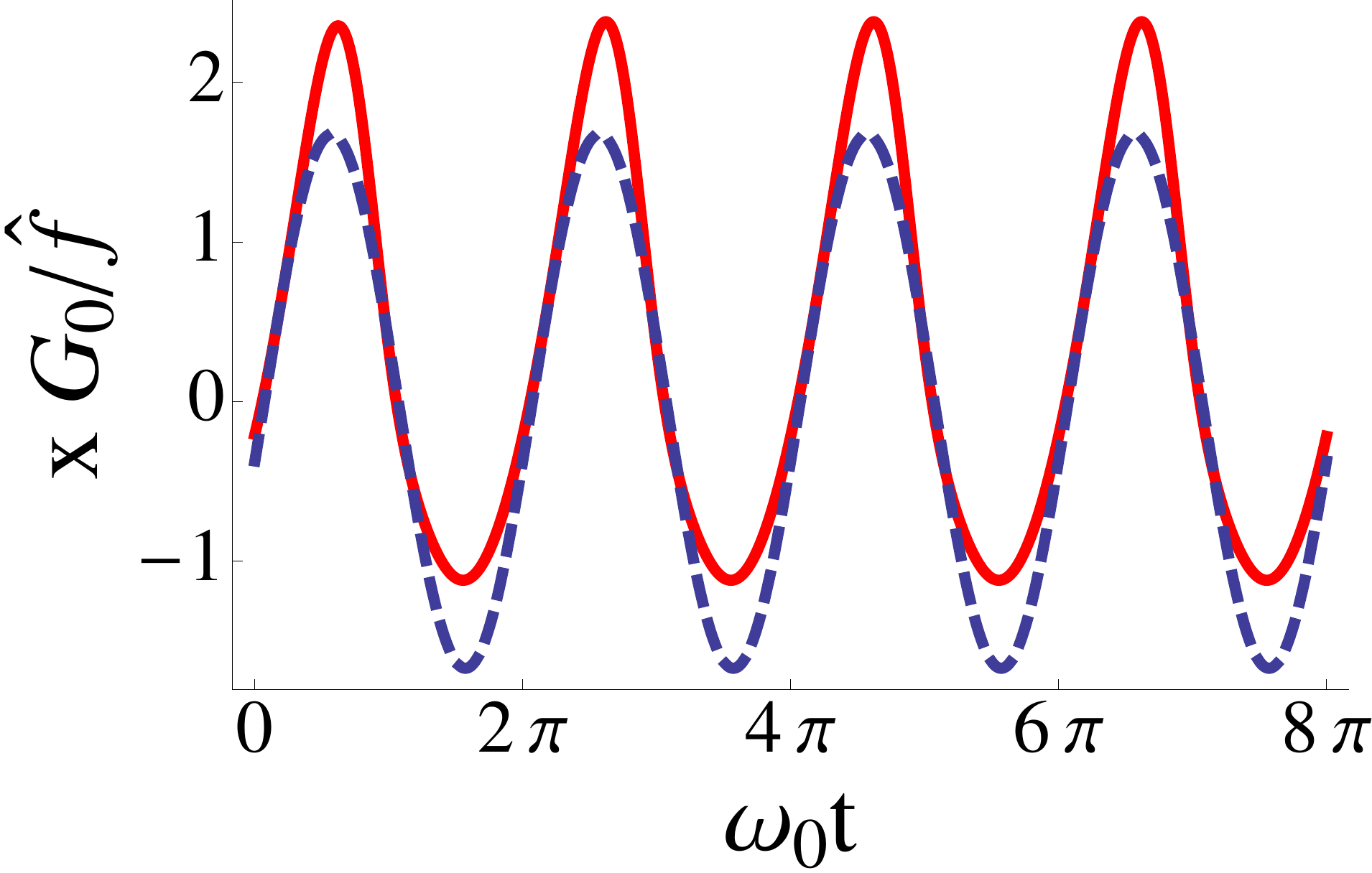} \\ [0.4cm]\\
\end{array}$
\end{center}
    \caption{Weakly nonlinear response of the inelastic power-law fluid (red
solid line) and the corresponding apparent linear response (blue dashed line),
reconstructed as described in the main text, displayed over driving force
(panel a) and time (panel b). Parameters are $\hat f \Delta x_b=0.35$,
$\delta=0.05$, $\omega\approx k$. Nonlinear effects in the rheology can already
be seen for much smaller amplitudes. }
    \label{fig:nonlin_example}
\end{figure}

This suggests that the interpretation for the peaked dissipative patterns
observed in Refs.~\citenum{Tharmann2007,Lieleg2008,Lieleg2009} in
terms of bond-breaking is not compatible with linear response. To see how
nonlinear effects can enter the {\em apparent} linear response, we note that it
is common practice to extract rheological spectra from a weakly nonlinear
response by Fourier-transforming the response and extracting the contribution in
resonance with the driving to reconstruct a fictitious sinusoidal response with
well-defined amplitude and phase shift (see figure~\ref{fig:nonlin_example} for
a somewhat exaggerated example). As nonlinear processes generally excite higher
harmonics but also change the resonance response, this practice is bound to
include a nonlinear bycatch.

In the following, we determine the nonlinear contributions to the apparent
linear susceptibilities by calculating the nonlinear response and isolating the
contributions in resonance with the driving frequency. In order to do this
in a formally fully consistent way, we would need to include higher-order terms
for the bond response and for the material response in the absence of bond
breaking. However, for our present purpose, which is to capture the essential
physics of the model rather than to deal with some circumstantial details, this
is in fact not necessary. We therefore assume favorable conditions, namely that
both the bonds and the material in the absence of bond breaking react linearly,
and that the only relevant source of nonlinearity lies in their mutual
interaction. Deviations from our results due to an actually nonlinear bond
response and due to potential mechanical nonlinearities in the employed
viscoleastic model (here the power-law fluid) can be evaluated numerically if
need arises. Their main effect will typically be a mere parameter
renormalization.

\begin{figure}[!h]
\begin{center}
$\begin{array}{c@{\hspace{3 mm}}c}
\multicolumn{1}{l}{\mbox{\bf\Large a}} &
	\multicolumn{1}{l}{\mbox{\bf\Large  b}} \\ [-0.43cm]\\
  \includegraphics[width=6 cm]{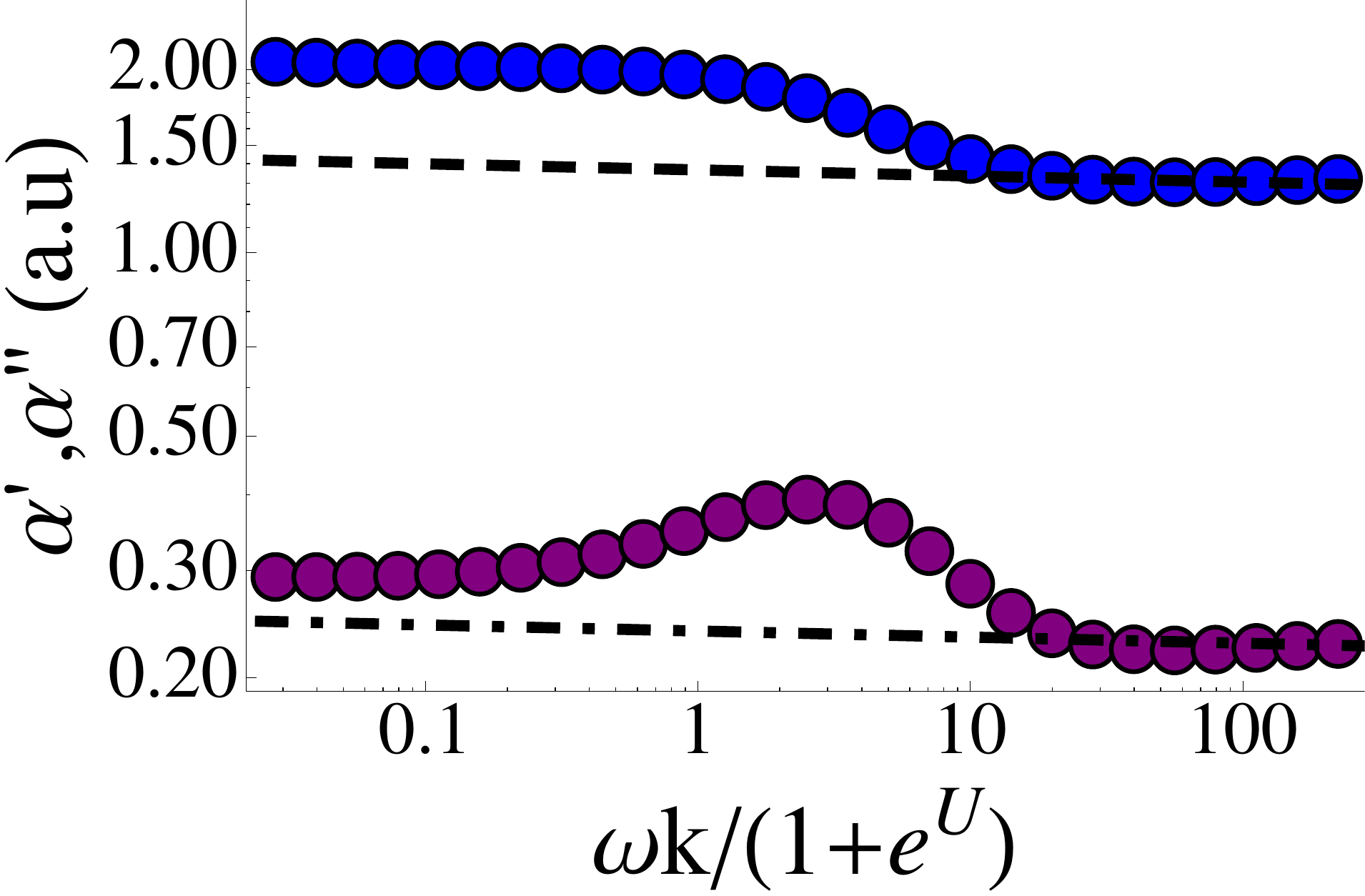} &
	\includegraphics[width=6 cm]{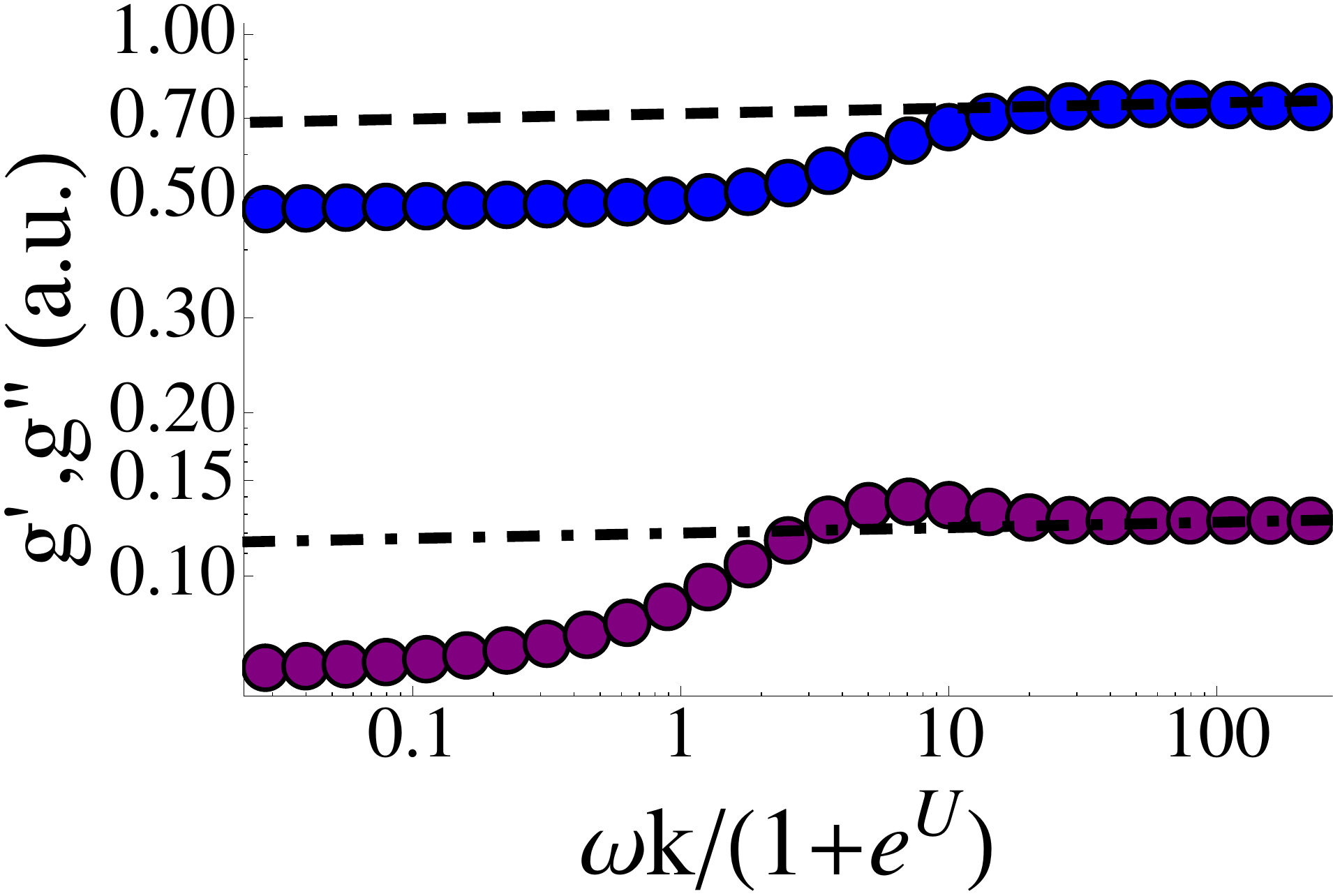} \\ [0.4cm]\\
% \mbox{\bf (aa)} & \mbox{\bf (bb)}
\end{array}$
\end{center}
    \caption{Nonlinear (symbols) and linearized (lines) susceptibilities (a) and
moduli (b) for the inelastic a power-law fluid defined in section
\ref{sec:model_power-law}. Due to the bond breaking, a shoulder develops in the
in-phase response (blue), and a peak is observed in the out-of-phase response
(purple). The parameters are $\delta=0.01$, $k=1.4\times 10^{-2}$, $U=1$, and
$\hat f\Delta x_b=0.35$.}
  \label{fig:nonlin_response}
\end{figure}

With these remarks in mind, we can use equation~(\ref{eq:response})
and expand it to higher order in $\hat f$. Up to third order and using the
abbreviation $d=\tan(\delta\pi/2 )$, the expansion reads
\begin{eqnarray}
    x(t)&\approx&\omega_0^{-\delta}\frac{\hat f}{G_0 \nu_0} 
\left(\sin(\omega_0 t)-d \cos(\omega_0
t)\right)\nonumber\\
  &+& G_0^{-1}\omega_0^{-\delta}\left(\frac{\hat
f}{\nu_0}\right)^2\left\lbrace\sin(\omega_0 t) -d\cos(\omega_0 t)
\right\rbrace\left\lbrace\alpha_\nu' \sin(\omega_0 t)-\alpha_\nu'' \cos(\omega_0
t)\right\rbrace \nonumber\\
&+&G_0^{-1} \omega_0^{-\delta}\left(\frac{\hat f}{\nu_0}\right)^3 
\left\lbrace\sin(\omega_0 t) -d \cos(\omega_0 t) \right\rbrace \left\lbrace
\alpha_\nu' \sin(\omega_0 t) -\alpha_\nu'' \cos(\omega_0 t)
\right\rbrace^2,\nonumber\\
\end{eqnarray}
with the equilibrium bond fraction $\nu_0=(1+e^{-U})^{-1}$.
By rewriting the trigonometric functions one can show that the terms of
order $\hat f^2$ only induce a response
at the first harmonic at $2 \omega_0$ and a constant offset. Therefore, the
first contribution to the apparent linear response comes from the third-order
terms. Dropping all terms that do not contribute to the response at
$\omega_0$, the apparent linear response becomes
\begin{eqnarray}
    x^{(3)}&\approx&\omega_0^{-\delta}\frac{\hat f}{G_0 \nu_0}
\left(\sin(\omega_0 t)-d \cos(\omega_0 t)\right)\nonumber\\
&+&G_0^{-1} \omega_0^{-\delta}\left(\frac{\hat f}{\nu_0}\right)^3 \frac{1}{4}
\times\nonumber\\
&\big\lbrace& \left[3 \alpha_\nu'(\omega_0)^2-\alpha_\nu''(\omega_0)^2-2 d
\alpha_\nu'(\omega_0)\alpha_\nu''(\omega_0) \right]\sin(\omega_0 t)\nonumber\\
&&-\left[d\alpha_\nu'(\omega_0)^2+3 d\alpha_\nu''(\omega_0)^2+2
\alpha_\nu'(\omega_0)\alpha_\nu''(\omega_0)\right] \cos(\omega_0 t)\big\rbrace.
\end{eqnarray}
The apparent linear susceptibility belonging to this response has the real and
imaginary parts
\begin{equation}
   \alpha^{(3)'} (\omega_0)= \frac{\omega_0^{-\delta}}{G_0 \nu_0}
\left\lbrace1-\frac1 4 \left(\frac{\hat
f}{\nu_0}\right)^2\left[\alpha_\nu''(\omega_0)^2+2 d
\alpha_\nu'(\omega_0)\alpha_\nu''(\omega_0)-3\alpha_\nu'(\omega_0)^2\right]
\right\rbrace
  \label{eq:response_3O_r}
\end{equation}
and
\begin{equation}
   \alpha^{(3)''}(\omega_0)= \frac{\omega_0^{-\delta}}{G_0 \nu_0}
\left\lbrace d+\frac1 4 \left(\frac{\hat
f}{\nu_0}\right)^2\left[3 d \alpha_\nu''(\omega_0)^2+2
\alpha_\nu'(\omega_0)\alpha_\nu''(\omega_0)+ d
\alpha_\nu'(\omega_0)^2\right]\right\rbrace
  \label{eq:response_3O_i}
\end{equation}
respectively (figure~\ref{fig:nonlin_response}). Obviously, we can isolate the
nonlinear bond-breaking contribution by
\begin{equation}
    \alpha_{\rm nl,bonds}(\omega_0)=\frac{\alpha^{(3)}(\omega_0)-\alpha^{(0)
}(\omega_0)}{\alpha^{(0)}(\omega_0)},
\label{eq:nonlin_response}
\end{equation}
where $\alpha$ is either $\alpha'$ or $\alpha''$ and $\alpha^{(0)}(\omega_0)$
stands for the linear susceptibilities of the material without bond breaking.

Equation~(\ref{eq:nonlin_response}) can be discussed analytically. By inspection
of equations (\ref{eq:response_3O_r})-(\ref{eq:nonlin_response}) we can see that
$\alpha^{(3)}$ exhibits the same scaling behavior as
equations~(\ref{eq:bonds_linear_re})-(\ref{eq:bonds_linear_im}). In particular,
the frequency scale is also set by $k(1+e^U)$. However, due to the nonlinear
combination of $\alpha_\nu$ with $\alpha_{\rm S}$ in
equation~(\ref{eq:nonlin_response}), the peak of the dissipative nonlinear
combined response, $\alpha^{(3)''}(\omega)$, will in general not be located at
$\omega=k(1+e^U)$. A simple calculation (see appendix) reveals that the
position of the peak is at
\begin{equation}
   \omega^*\approx\frac 1 {\sqrt 3} k \cdot (1+e^{U}).
  \label{eq:omegastar}
\end{equation}
\begin{figure}[h]
\begin{center}
$\begin{array}{c@{\hspace{3 mm}}c}
\multicolumn{1}{l}{\mbox{\bf\Large a}} &
	\multicolumn{1}{l}{\mbox{\bf\Large  b}} \\ [-0.43cm]\\
  \includegraphics[width=6 cm]{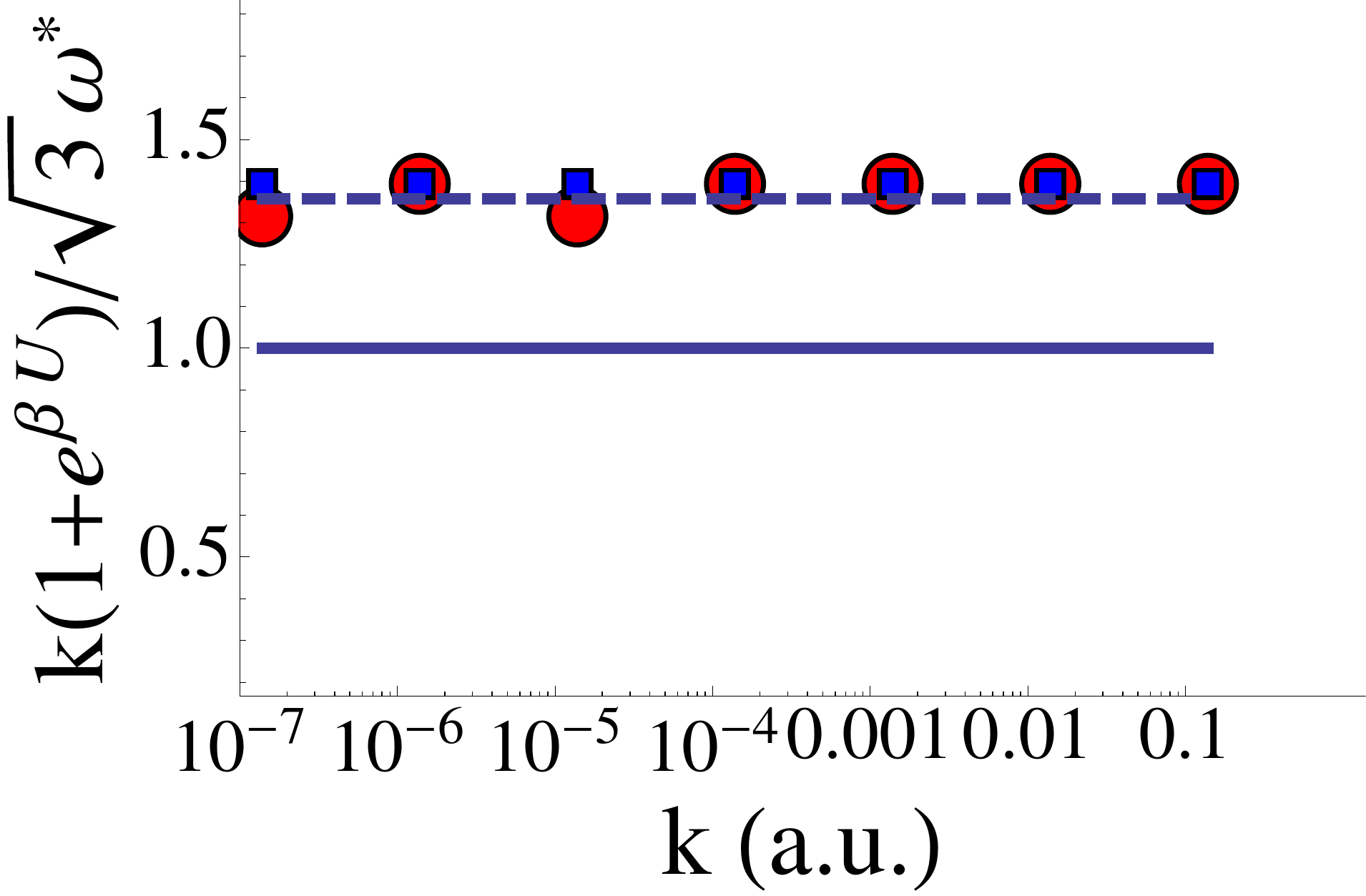} &
	\includegraphics[width=6 cm]{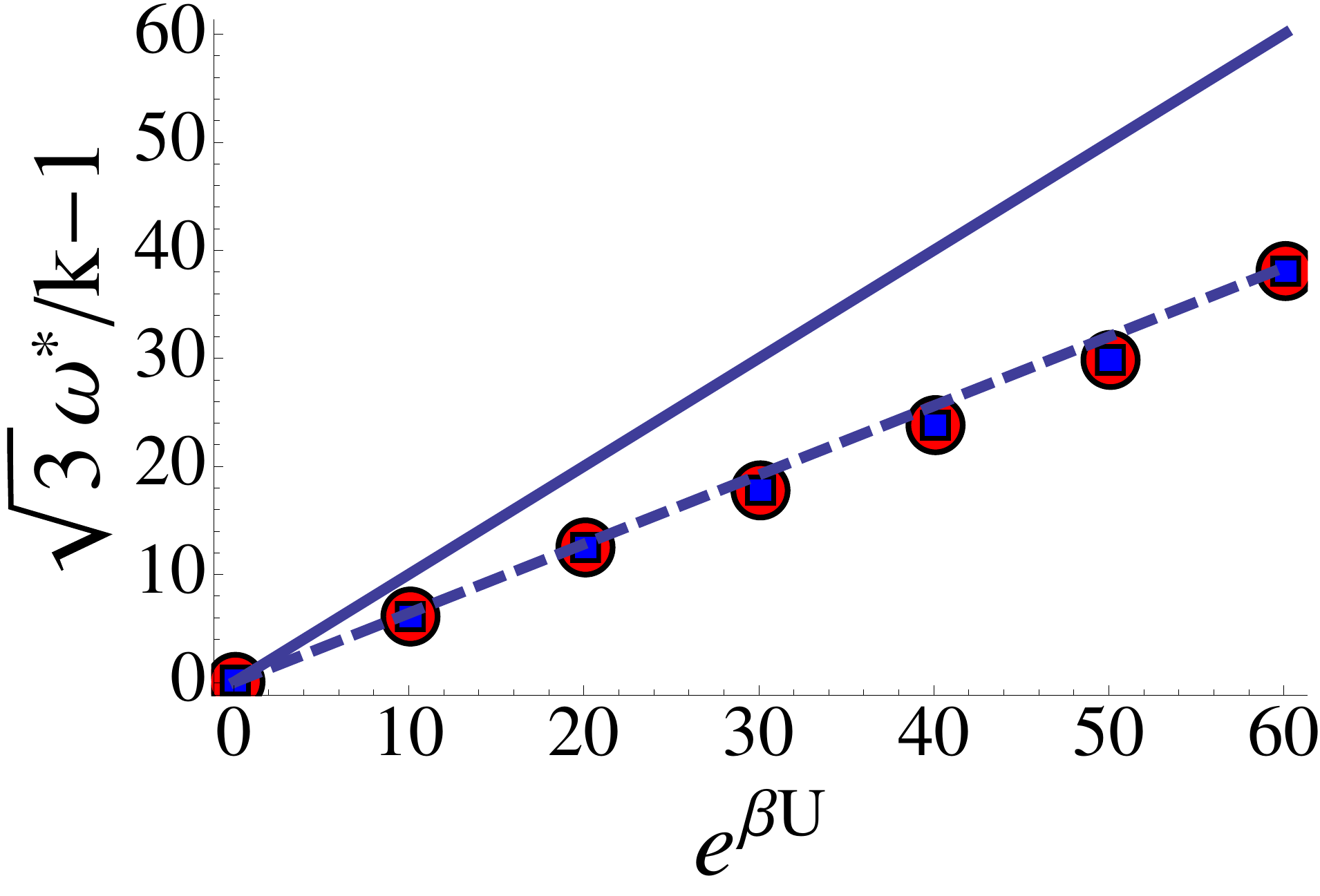} \\ [0.4cm]\\
% \mbox{\bf (aa)} & \mbox{\bf (bb)}
\end{array}$
\end{center}
    \caption{Position of the peak in the imaginary part of the nonlinear
susceptibility. Numerical evaluations for the power-law fluid (blue
squares) compared to numerical evaluations for the inelastic GWLC (red circles,
see section~\ref{sec:results_GWLC}) and the analytical prediction,
equation~(\ref{eq:omegastar}) (solid lines) and the semi-phenomenological
correction, equation~(\ref{eq:omegastar_ext}) (dashed
lines, see main text), for various values of the equilibrium off
rate $k$ (panel a, $10^{-8}<k<0.1$) and the equilibrium relative binding
affinity $e^{U}$ (panel b, $0<e^{U}\leq 64$). The driving
amplitude was $\hat f \Delta x_b=0.12$, other parameters as in
figure~\ref{fig:nonlin_response}. For the numerical Euler scheme, the time step
was chosen as $(2 \pi)^{-1} \omega_0 \Delta t =5\times 10^{-5}$.}
  \label{fig:comp_power}
\end{figure}

This approximate nonlinear formula gives us an analytical prediction (without
any free parameter) of the position of the peak in the spectrum in dependence of
the bond parameters. To assess the quality of the analytical prediction for the
peak position, we numerically evaluate
equations~(\ref{eq:kon})-(\ref{eq:driving_force}) (using an Euler scheme)
together with equation~(\ref{eq:alphaSdoubleprime}) (an example is provided in
figure~\ref{fig:nonlin_response}) and extract the peak frequency for
various values of either $k$ or $U$, keeping all other parameters constant. The
results are presented in figure~\ref{fig:comp_power}. We find that the
functional dependence is predicted correctly, but that the theoretical
predictions are off by a constant factor of about $0.6$ (figure
\ref{fig:comp_power}b). This is so because we assumed the bonds to always
respond linearly and neglected the second-order contribution to the bond
response, which actually also contributes to the third order of the displacement
response. If one is interested in quantitative parameter estimations, it is
straightforward to take into account the second-order contribution. However, the
more extended formula is rather cumbersome. In practical applications,
additional nonlinear corrections are more conveniently taken into account by a
phenomenological correcting factor. Multiplying $e^U$ on the right-hand side
of equation~(\ref{eq:omegastar}) by
$0.64$,
\begin{equation}
   \omega^*\approx\frac 1 {\sqrt 3} k \cdot (1+0.64 e^{U}),
  \label{eq:omegastar_ext}
\end{equation}
leads to a good quantitative agreement with the numerical results
for a broad range of $k$ and $e^U$, (figure~\ref{fig:comp_power}, dashed lines),
independently of the driving force amplitude $\hat f$ for $\hat
f \Delta x_b\lesssim 0.18$ (see supplementary figure 1).

The amplitude of the absorption peak turns out to be much more sensitive to the
intrinsic nonlinearity of the bond dynamics. Deriving a simple approximate
formula as for the peak position is therefore much less useful. Nevertheless, we
can predict the approximate dependence on some key parameters from the more
complicated, fully consistent
expressions.  Here we only mention that the amplitude is independent of the
equilibrium off rate $k$, depends on the relative affinity as $(1+e^U)^{-1}$
and depends quadratically on the driving force amplitude $\hat f$ (data not
shown). We further note that it is easy to show from the analytic equations
that asymptotically, for low frequencies, both the in-phase and the out-of-phase
part of the susceptibility are merely shifted from the linear-response to higher
values, where the shift is stronger for the in-phase susceptibility (see
figure~\ref{fig:nonlin_response}). The interpretation is that the network softens, while the dissipation
increases due to forced bond-breaking. Note that the high-frequency response is
not affected by bond breaking, while the low-frequency power-law asymptotics
violates the Kramers-Kronig relations, testifying the nonlinear character of the
modulus.

\subsection{Apparent linear response in the inelastic GWLC model}
\label{sec:results_GWLC}

The inelastic power-law fluid analyzed in the preceding section is to be
understood as a schematic model for weakly crosslinked biopolymer networks. A
more realistic but still minimalistic description is given by the recently
proposed model of an inelastic glassy wormlike chain \cite{Wolff2010} that
builds on the linear modulus of the phenomenologically highly successful glassy
wormlike chain (GWLC) model \cite{Semmrich2007,Kroy2007,Glaser2008} and combines
it with a bond kinetics scheme as outlined in section~\ref{sec:model_GWLC}.
Due to the slightly more complex structure of the GWLC modulus compared to the
power-law fluid, we resort to numerical evaluations to explicitly work out the
response. We can again compare these numerical results to the analytical
predictions obtained above to better assess the plausible range of validity of
our analytical expressions in applications to biopolymer networks.

\begin{figure}[h]\begin{center}
		      \includegraphics[width=6 cm]{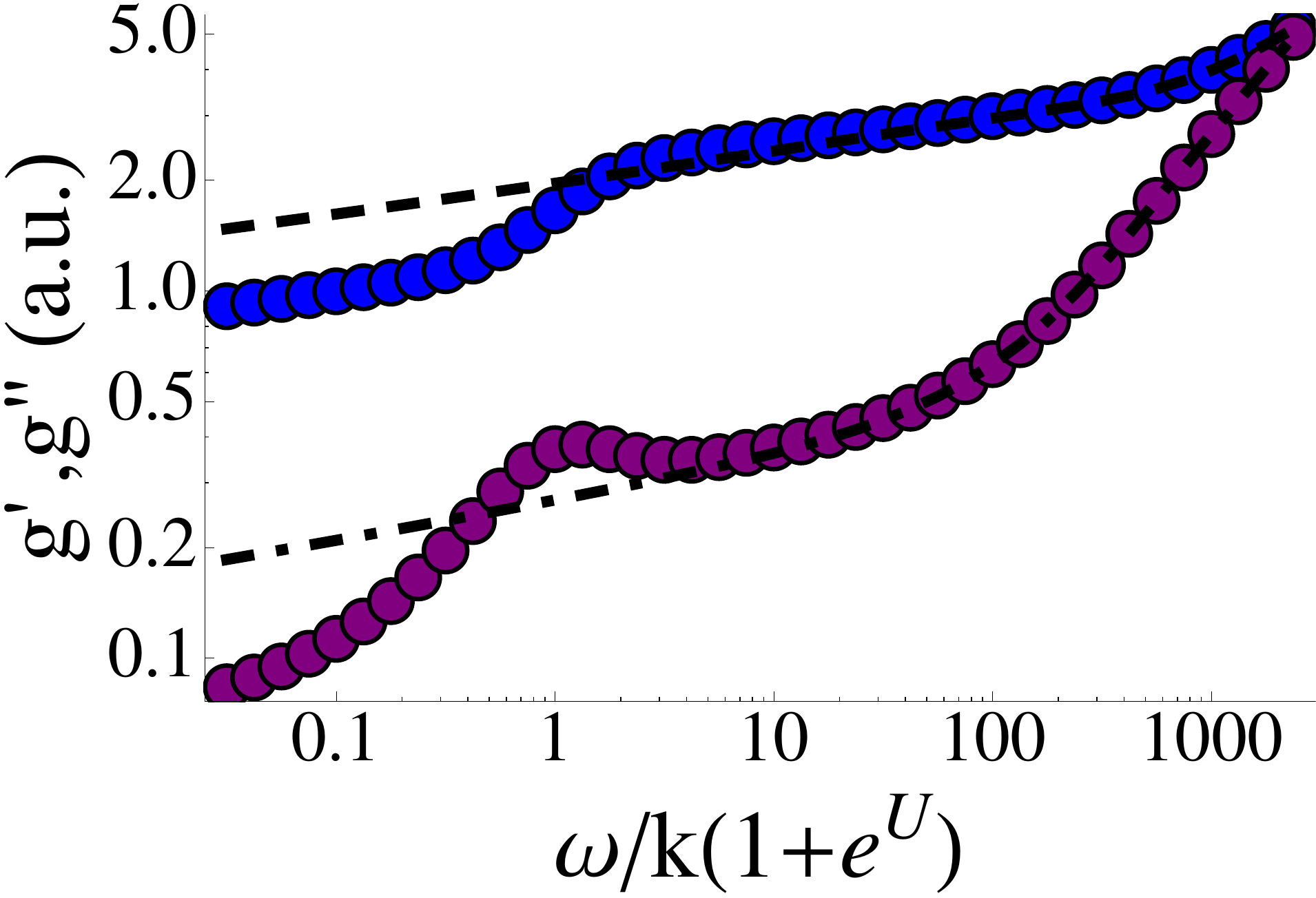}
                 \end{center}
  \caption{Apparent linear storage (blue) and loss (purple) modulus {\it
versus} frequency (arbitrary units). The lines represent the strictly linear
viscoelastic storage and loss modulus in the absence of bond breaking.
The parameters are $k=1.4\times 10^{-2}$, $U=-1$, and $\hat f\Delta x_b=0.15$. }
  \label{fig:nonmon_mods}
\end{figure}

As established above for the inelastic power-law fluid, bond breaking matters
only for the nonlinear response. We therefore again operationally define a set
of apparent linear moduli by isolating the resonance with the driving frequency
in the Fourier spectrum of the response. Evaluating the nonlinear response of
the inelastic GWLC as described in Ref.~\citenum{Wolff2010} and calculating the
apparent linear moduli for oscillatory stress control, we find a qualitative
agreement with measurements on crosslinked actin networks
\cite{Tharmann2007,Lieleg2008,Lieleg2009,Broedersz2010}. In particular, we find
again a maximum in the loss modulus and a shoulder in the storage modulus as
illustrated in figure \ref{fig:nonmon_mods}.

\begin{figure}[h!]
\begin{center}
$\begin{array}{c@{\hspace{3 mm}}c}
\multicolumn{1}{l}{\mbox{\bf\Large a}} &
	\multicolumn{1}{l}{\mbox{\bf\Large  b}} \\ [-0.43cm]\\
  \includegraphics[width=6 cm]{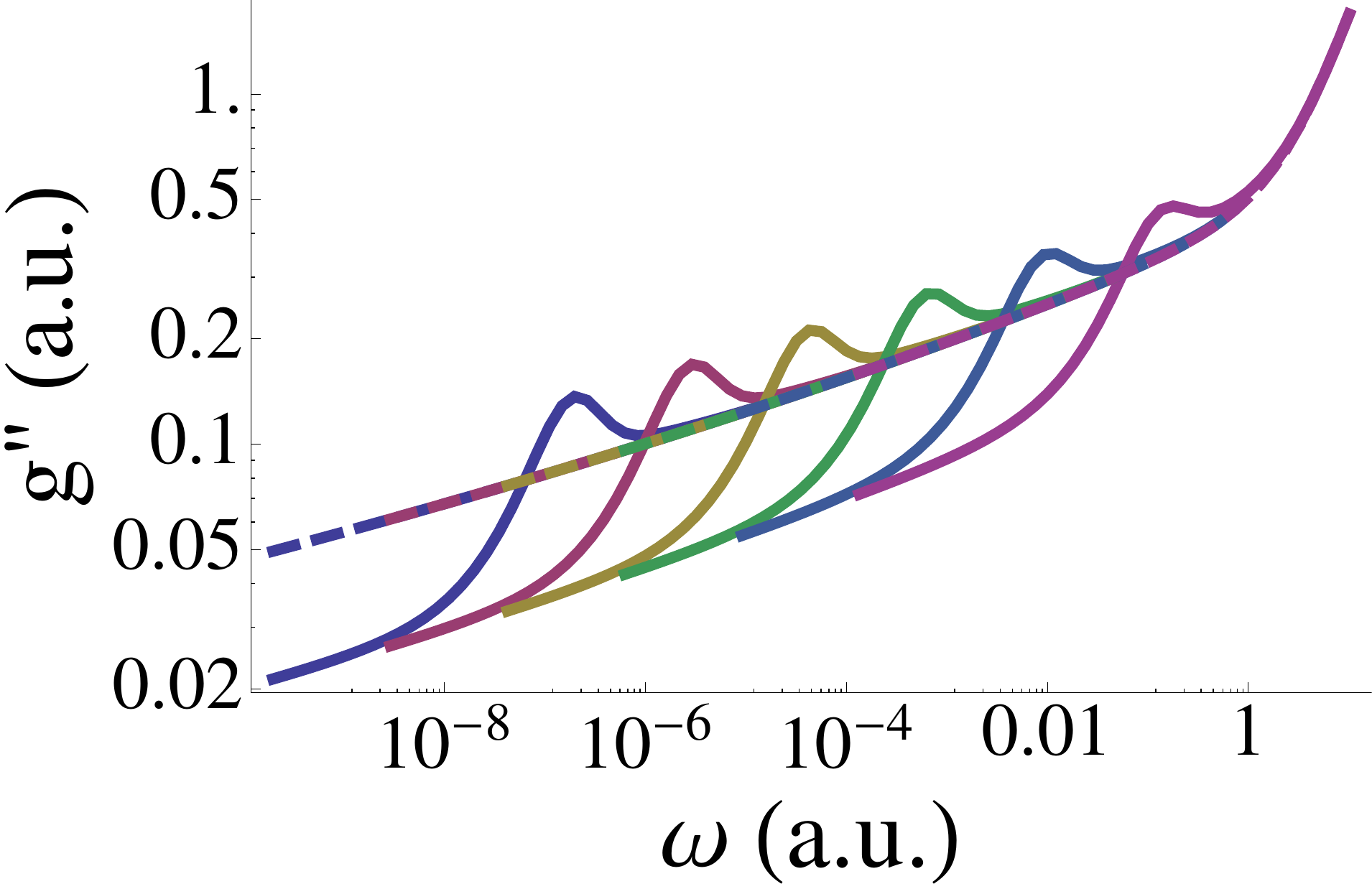} &
	\includegraphics[width=6 cm]{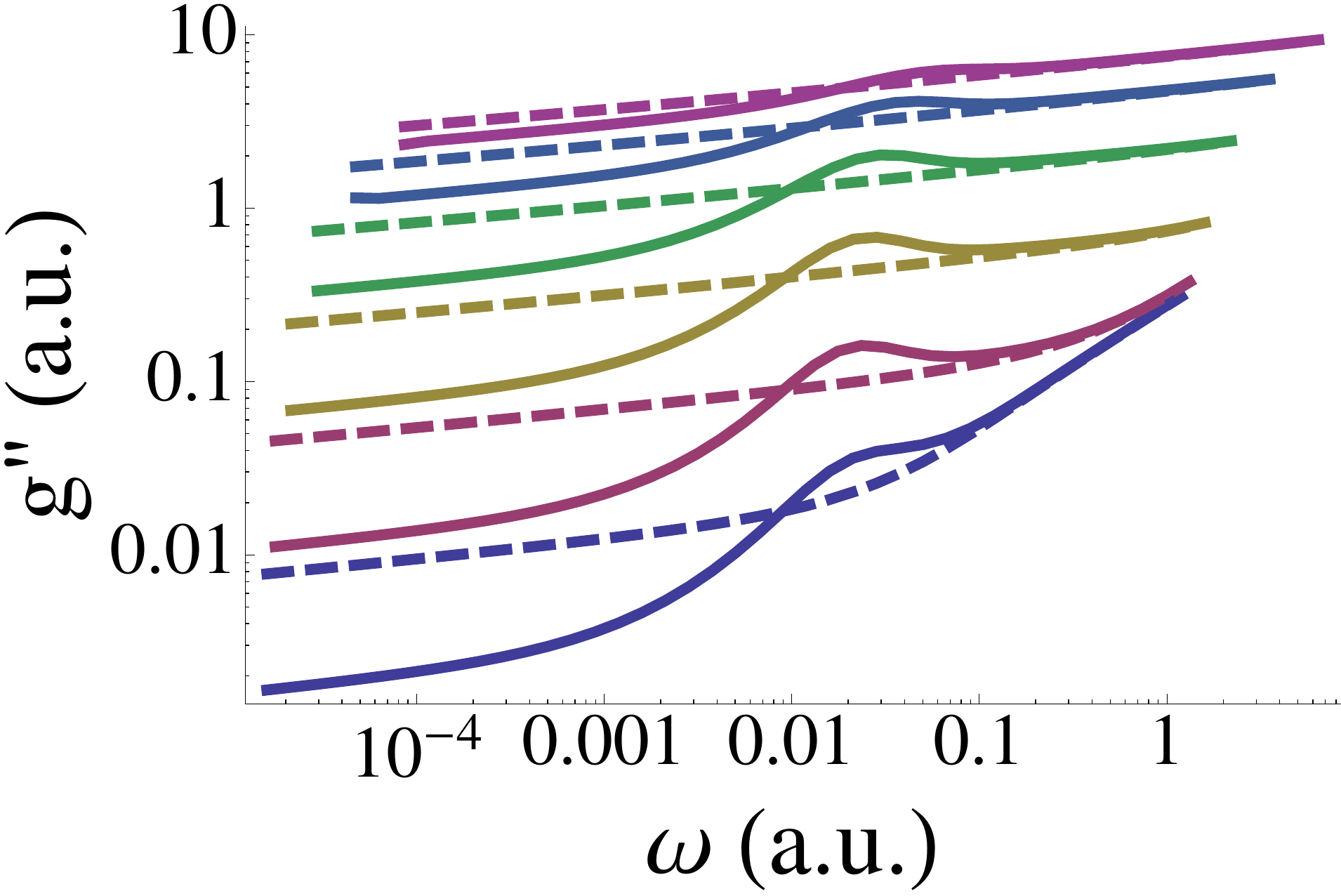} \\ [0.4cm]\\
\end{array}$
\end{center}
    \caption{Apparent linear rheology of an inelastic GWLC. Frequency-dependent
loss modulus for various values of $k$ (panel a, $10^{-8}<k<0.1$, increasing
from left to right) and for various values of the relative binding affinity
$e^U$ (panel b, $0<e^{U}\leq 10$, increasing from bottom to top), all other
parameters as in figure~\ref{fig:nonmon_mods}.}
  \label{fig:parameter_dependence}
\end{figure}

It is apparent from figure \ref{fig:nonmon_mods} that the frequency dependence
of the linear moduli superimposes with the bond effects in a first
approximation. Therefore, we can easily isolate the nonlinear part of the
response (see section~\ref{sec:results_power-law}). To numerically extract the
parameter dependence, we evaluate the nonlinear frequency-dependent response of
the inelastic GWLC at various values of either $k$ or $U$, keeping all other
parameters constant (figures~\ref{fig:comp_power} and
\ref{fig:parameter_dependence}). There are essentially two routes to change $k$
(see sections \ref{sec:model_bonds} and \ref{sec:model_GWLC}), one assuming that
changing $k$ also affects the linear modulus of the GWLC (by changing the
barrier height $\mathcal E$) and one that leaves the linear modulus unaffected
(assuming that only the microscopic attempt rate $\tau_0^{-1}$ or the bond
stiffness changes). Numerical evaluations show that both routes lead to the same
results for the peak positions within numerical tolerance (see supplementary
figure 2). In contrast, changing $U$ always influences the linear modulus,
because $U$ affects equilibrium properties such as the steady-state bond
fraction. 

The numerical evaluation of the inelastic GWLC in figure
\ref{fig:parameter_dependence} clearly reveals the dependence of the peak
position on the parameters. In the case of a strongly frequency-dependent
or very large modulus (fig \ref{fig:parameter_dependence}b), the
dissipation due to bond-breaking may be insufficient to become clearly
discernible in the rheological spectra, but it can always be uncovered by the
subtraction scheme given in equation~(\ref{eq:nonlin_response}). 

The actual position of the peak in dependence on $k$ (figure
\ref{fig:comp_power} a) and $e^{U}$ (figure~\ref{fig:comp_power} b) obeys
the  prediction, equation~(\ref{eq:omegastar_ext}), from the analytical
calculations in section~\ref{sec:results_power-law} very well (note that we do
not have any free parameters to adjust). Most strikingly, the numerical values
for $\omega^*$ for the inelastic GWLC agree within the numerical error with the
numerical values for the inelastic power-law fluid, indicating that the
dissipative pattern is reasonably robust. Its characteristic properties are
dominated by the generic bond breaking mechanism and relatively insensitive to
details of the rheology of the uncrosslinked polymer solution. 

\subsection{Exemplary application to actin/$\alpha$-actinin networks}

\label{sec:results_parameters}

Our analytical prediction for the position of the peak in
equation~(\ref{eq:omegastar_ext}) is sufficient to extract valuable kinetic
information from bulk rheological measurements. While, in principle, one could
also exploit the information encoded in the peak amplitude, this would require a
more sophisticated fitting procedure taking into account detailed experimental
information on the nonlinearity of the response. As this information in not
available for pertinent published data, we restrict the following analysis to an
interpretation of the peak position, which depends on the equilibrium off rate
and the relative affinity, with equation~(\ref{eq:omegastar_ext}). To exemplify
the procedure, we use data for an {\it in vitro }reconstituted
actin/$\alpha$-actinin network\cite{Broedersz2010}, where a peak in the apparent
linear loss modulus was observed.

Assuming power-law rheology to be a valid model for the
(strictly) linear response, we first use the high-frequency data for $g'$ to
estimate $G_0 \nu_0$ and $\delta$ to $4.7$ and $0.11$, respectively. Next, we
use equation~(\ref{eq:alphaSdoubleprime}) to
calculate the linear loss modulus $\alpha^{(0)''}$. Then, we use
equation~(\ref{eq:nonlin_response}) to isolate the peak, which turns out to be
located at $\omega^*\approx 2 s^{-1}$.  Using an affinity value from the
biochemical literature \cite{Goldmann1993}, $K=2.5\times10^{6} M^{-1}$, together
with actin and $\alpha$-actinin concentrations of $c_A=23.8 \mu M$ and
$c_\alpha= 0.238$, respectively \cite{Broedersz2010}, we obtain $e^U=K
(c_A+c_\alpha)\approx 60$.

From equation~(\ref{eq:omegastar_ext}), we find an equilibrium off rate of
$k\approx0.09 s^{-1}$, which agrees with the value obtained for
$\alpha$-actinin/actin in mechanical single-molecule
experiments \cite{Ferrer2008} within the experimental uncertainty. Since our
analysis is based on a simple (Bell-type \cite{Bell1978a}) bond-breaking model
the numerical agreement should not be over-interpreted, but is certainly
reassuring. Note that the rate obtained via either route is a factor of five
smaller than values obtained biochemically \cite{Goldmann1993}. It has been
argued that this discrepancy could be due to the restriction of the reaction
coordinate by directional loads in single-molecule experiments
\cite{Ferrer2008}. However, we noted an interesting fact that may or may not be
a systematic feature. If we follow experimental hints from actin/HMM solutions
that $\omega^*$ is approximately independent of actin- and crosslinker
concentration \cite{Lieleg2009} and consider the concentration values of $c_A=2
\mu M$ and $c_\alpha= 3 \mu M$ used in the biochemical bulk measurement
\cite{Goldmann1993}, we obtain $k\approx 0.4 s^{-1}$, which is exactly the value
reported. This would suggest that the equilibrium off rate depends on either the
actin- or the crosslink concentration. Another possible explanation could be
that different types of $\alpha$-actinin have been used or that biochemical and
biomechanical rate measurements systematically differ by a factor, which cancels
out when taking the ratio of the rates to calculate the affinity. This problem
certainly deserves a dedicated experimental investigation testing the various
hypotheses and possibly involving more elaborate models of bond breaking
\cite{Tshiprut2008, Freund2009}, which is beyond the scope of the present
contribution.

\section{Conclusions}

We have put forward a simple mechanistic model for the weakly nonlinear rheology
of reversibly crosslinked biopolymer networks. We verified our analytical
predictions against numerical results from two generic models for the
viscoelastic response of biopolymer solutions and networks and found good
agreement. Thereby, we established the characteristic dissipative pattern due
to bond breaking as a robust and reproducible rheological signature of
transiently crosslinked biopolymer solutions with well-defined crossliker
kinetics. The model seems suitable to explain some puzzling features in the
frequency-dependent rheology of crosslinked actin networks in terms of the
underlying physics and demonstrate the non-Maxwellian nature of what could
easily be mistaken for a classical terminal relaxation pattern in a limited
frequency interval.

There are two main conclusions to be drawn from the results presented above.
First, the non-trivial rheology of transiently crosslinked biopolymer networks
as observed in Refs.~\citenum{Lieleg2008,Lieleg2009,Broedersz2010} can also be
found in a conceptually very simple model of an inelastic biopolymer model, the
inelastic GWLC. This supports the expectation that the inelastic GWLC is indeed
a suitable minimal model for weakly crosslinked biopolymer networks. Second, as
the frequency-dependence of the maximum in the loss modulus is very well
predicted by our perturbation theory, it seems plausible that the mechanism
put forward in section~\ref{sec:results_power-law} is suitable for deducing
information on the crosslinker kinetics from the rheological spectra. In doing
so, one implicitly acknowledges the intrinsically nonlinear character of the
absorption peaks revealed by our analysis. 

Compared to single-molecule assays, which provide an alternative method to
probe crosslinker properties \cite{Ferrer2008}, rheology has the advantage of
being an ensemble method. A potential caveat of the outlined method is that the
procedure to extract the kinetic information from the modulus relies on the
knowledge of the linear modulus in the absence of bond breaking (see
section~\ref{sec:results_power-law}). This information may sometimes be
experimentally inaccessible, {\it e.g.\ }if true linear response conditions are
technically difficult to realize. However, we discussed the nonlinear rheology
for two generic models for the linear viscoelastic response of biopolymer
solutions and networks, and we anticipate that for most practical applications
at least one of them will provide a reasonable approximation.

The results presented in this paper thus help to turn a widespread speculation
about the physical mechanism providing biopolymer networks (and possibly also
living cells) with their extraordinary mechanical properties into a concise
quantitative prediction amenable to detailed experimental investigation.

\section*{Acknowledgements}

We thank A.~Kramer for preparing figure~\ref{fig:cartoon}, S.~Sturm for
designing the graphical abstract, and C.-M.~Pop, E.~Frey, and
J.~Glaser for helpful discussions and acknowledge financial support by the
graduate school ``Leipzig School of Natural Sciences --
BuildMoNa'' within the German excellence initiative.

\begin{appendix}
    
\section{Peak position in the nonlinear response}

Using the abbreviations $x=\omega/(1+e^{U})k$, $h(x)=\alpha_{\rm
nl, bonds}''$, $b=((\Delta x_b+\Delta x_u)/(4 \cosh^2(U/2))\cdot\hat
f/ \nu_0)^2$, and $d=\pi/2 \tan(\delta)$,
equation~(\ref{eq:nonlin_response}) becomes
\begin{equation}
    h(x)=\frac{3 b d}{4} \left(\frac{x}{1+x^2}\right)^2+\frac b 2 \frac
x{(1+x^2)^2}+\frac{b d} 4\frac 1{(1+x^2)^2}.
\label{eq:app_h}
\end{equation}

The peak position $x^*$ is determined by the condition 
\begin{equation}
    \frac{d}{dx} h(x) \bigg \vert_{x=x^*}=0.
\label{eq:app_cond}
\end{equation}

After some simple algebra, equations~(\ref{eq:app_h}) and (\ref{eq:app_cond})
yield the condition
\begin{equation}
    x^{*3}+\frac{x^{*2}}d-\frac 1 3 x^*-\frac 1{3d}=0.
\end{equation}

For $d>0$, this equation has only one positive real solution, $x^*=1/\sqrt{3}$.
Noting that $x^*=\omega^*/(1+e^{U})k$, we obtain equation~(\ref{eq:omegastar})
for the frequency of the peak.

\end{appendix}

\bibliographystyle{rsc}
\bibliography{library}

\end{document}